\documentclass[12pt]{article}

\usepackage{graphicx}

\begin{document} 

\vskip1cm

\begin{center}  {\Large {\bf Effects of Minijets on Common Observables in Heavy-Ion Collisions with Uncommon Implications}}
\vskip .75cm
 {\bf Rudolph C. Hwa}
\vskip.5cm
{Institute of Theoretical Science and Department of
Physics\\ University of Oregon, Eugene, OR 97403-5203, USA\\}
\vskip.5cm
{\today}
\end{center}

\vskip.5cm
\begin{abstract} 
\vskip.5cm
In this brief review of the observable effects of minijets in heavy-ion collisions the main points emphasized are that the quadruple moment $v_2(p_T,b)$ and the hadronic ($\pi$ and $p$) spectra  at low $p_T$ can both be reproduced by minijet contributions to the recombination of thermal and shower partons. Without using hydrodynamics the minijet approach does not trace the evolution of the expanding system. The thermal distribution of the medium partons at the time of hadronization is assumed, but rapid thermalization initially is not required so as to allow minijets to leave their footprints on the system in the final state. Azimuthal anisotropy due to minijets is directly calculated in the momentum space without any fluid assumption relating the spatial eccentricity to $v_2$. There are no more parameters used, compared to the hydro approach in fitting the data on $v_2$ and $p_T$ spectra. Thus both approaches satisfy the sufficiency condition for a viable description of the dynamical process involved.

\end{abstract}
\vskip0.5cm

\section{\large Introduction}
Theoretical descriptions of the physics of relativistic heavy-ion collisions can mostly be partitioned into two camps, as is conventionally done, without any contradiction with each other because they refer to two different kinematic regions well separated by the transverse momentum $p_T$ \cite{hw}.  At $p_T < 2$ GeV/c hydrodynamics has been regarded as the standard model \cite{nsac}, while at $p_T > 6$ GeV/c perturbative QCD is broadly recognized as the relevant theory to treat the hard-scattering dynamical processes \cite{hw}.  There are numerous other problems, such as phase transition, quarkonium and heavy-flavor production, gluon saturation, etc., that all have significant places that intersect with the dual theme of a collective dense medium on the one hand and the microscopic processes of parton interaction on the other.  While this grand picture is a useful summary of the achievements in the program of studying hot QCD matter in heavy-ion collisions, there are areas worthy of internal debate within the community so that critical physics issues are not covered up for the sake of presenting a unified picture to the outside.

In this brief review we focus on one topic only:  the effects of minijets at low $p_T$.  The subject matter does not belong to any of the conventional domains mentioned above.  Minijets are hard to define precisely because they are jets produced in the intermediate-$p_T$ region for which pQCD is not reliable.  Their effects in the low-$p_T$ region compete with the results of hydrodynamics, and therefore can spoil the polished picture that the standard model presents.  But in order for a model to become standard, it is necessary to rule out all alternative possibilities.  Since no one can prove the necessity of any theory, all that one can do is to present phenomenological evidences that certain approaches contain features worthy of consideration as a possible explanation of the data, thus satisfying the sufficiency condition of  a viable model.  Hydrodynamics is one such approach.  Minijet production is another possible approach.

\section{\large The Hydro and Minijet Approaches}
The hydro approach is based on relativistic hydrodynamics supplemented by a number of inputs that include initial geometry, equilibration time, equation of state, viscosity, hadronization scheme, etc. \cite{kh,te}.  Credibility for the approach has been drawn from fitting the data on the second harmonics of the azimuthal anisotropy, $v_2$, usually referred to as elliptic flow.  The equilibration time $\tau_0$ is taken to be 0.6 fm/c in most calculations, and hadronization is by means of Cooper-Frey's freeze-out prescription, which assumes a sudden transition from a fluid in local thermal equilibrium to free-streaming hadrons \cite{cf}.  Viscosity, though originally neglected, was later incorporated in viscous hydrodynamics \cite{te}.  Fluctuations of the initial configuration has also been considered to account for the features seen in the data on $\phi$ anisotropy and their higher moments $v_n$ \cite{alver}.  Each step along the way more precise data put constraints on the theory that improve the agreement with observation.  Indeed, the creation of a strongly-interacting quark-gluon plasma constitutes a satisfactory picture that can be presented as the result of the enormous effort spent on the heavy-ion program.

There are, however, scruples that one can raise on whether that is a complete picture.  For an extended object local equilibration need not be accomplished at a universal time $\tau_0$.  Since the region in the middle of the overlap has higher density than on the edges, thermalization may take different time durations.  Hard and semi-hard partons produced within 1 fm from the boundary can emerge from the medium before the interior is fully thermalized.  If the $p_T$ of those jets are less than 5 GeV/c, there can be many of them.  Their effects are not accounted for in hydrodynamics.  Those minijets can lead to azimuthal anisotropy in non-central collisions, since the hadron distributions correlated to triggers are known to depend on the trigger direction \cite{af}.  Furthermore, the ridge phenomenon is associated with jets, whether or not a trigger is used \cite{jp,md,tk}.  Although triangular eccentricity has been suggested as the source of some $\phi$ characteristics \cite{ar}, including ridges, one should be open to the possibility that minijets can contribute to fluctuations of the initial configuration and may well be the origin of some high-order eccentricity moments.

The various issues raised above remain at the level of questions on the completeness of the hydro approach until appropriate phenomenology can relate the effects of minijets to observations --- especially on the data that have given support to hydro calculations.  In various respects that has already been done, as in Refs. \cite{cw, tt, rh}.  To recognize the importance of minijets does not mean that one must abandon the notion of expansion of the dense medium created in heavy-ion collisions.  It suggests that the expansion is more complicated than can be described by hydrodynamics alone in the way so far applied.  To open up the possibility that it may be inadequate, it is necessary to show the relevance of other effects, even if they are by themselves also incomplete.  We summarize in the following sections the minijet approach \cite{hz} formulated in the framework of the recombination model \cite{hy, hy1}.

\section{\large Ridge, Minijets and Azimuthal Anisotropy}
To relate characteristics in the $\eta$ and $\phi$ variables of the produced hadrons to eccentricities of the initial configurations assumes the reliability of the theoretical description of the evolutionary process leading from the initial to final states.  If the adequacy of that description is questioned, it is necessary to approach the subject of ridges and $v_2$ in an independent way.  There is a large body of experimental evidences that related ridges to minijets \cite{ja,ba,aver,ba1}.  With such correlation data in mind we move a step further by advancing the idea that the ridge particles should be a non-negligible part of the single-particle spectra, even when there is no trigger as used in correlation experiments.  A minijet can give rise to ridge particles whether or not the minijet itself is detected.

If a semihard parton created near the surface is directed inward, it would be absorbed by the medium and become thermalized.  Its recoil parton directed outward can get out of the medium after losing some momentum.  That minijet should lead to observable consequences in the hadron spectra.  The hadronization process is more complicated than that described in the Cooper-Frye scheme, since the jet component is not in thermal equilibrium with its immediate environment.  The semihard parton that emerges fragments to shower partons (S), which in turn can recombine with the thermal partons (T) of the medium to form a variety of hadrons, the strongest component of which being the pion.  Hadronization occurs at late time, so thermal partons have time to equilibrate locally, but need not follow a hydro prescription that requires rapid thermalization at early time.  The whole system can expand in ways that cannot be described by a universal formula from beginning to end.  The energy loss of a semihard parton on its way out of the medium can lead to enhancement of the thermal motion of the medium partons in the vicinity of the trajectory.  If an exponential distribution is used to represent the thermal partons at the end, then the inverse slope $T$ near the semihard parton can be higher than that of the background.  We shall identify the pions formed from those enhanced thermal partons by TT recombination as the Ridge particles, while the pions from TS recombination build up the peak (that sits above the ridge), referred to as Jet \cite{jp}.  Both components would be absent without the semihard parton that initiates the Ridge (R) and the Jet (J), which are $\phi$ dependent.  The background thermal partons recombine to form the Base (B), which is $\phi$ independent.  The separation of the hadron spectra into these components is the starting point of our approach in the absence of a calculational scheme to trace the evolution of the system from collision to hadronization.  Our task is to show that such a starting point can lead to results that are in good agreement with all relevant low- to intermediate-$p_T$ data on common observables without using any more free parameters than in any other approach.

To be explicit, let us write the single-particle distribution $\rho^h(p_T,\phi,b)$ for hadron $h$ at mid-rapidity $\eta \approx 0$, where $b$ denotes impact parameter, in the form 
\begin{eqnarray}
\rho^h(p_T,\phi,b) =B^h(p_T,b) + R^h(p_T,\phi,b),     \label{1}
\end{eqnarray}
for $p_T < 2$ GeV/c.  At this point we omit mentioning the J component (by TS recombination) for simplicity.  We put aside the consideration of the $\eta$ dependence for now and focus on the $\phi$ dependence.  The aim is to reproduce the data on $v_2(p_T)$ without using the hydro concept of flow and pressure gradient.  The crucial step to take is to formulate a way that can relate the spatial anisotropy of the initial configuration to the momentum anisotropy of the final state, based on the physics that semihard partons are the initiators of the anisotropy.  That relationship was first obtained in a naive treatment of the initial configuration \cite{rch,chy}.  A more thorough study that includes the azimuthal correlation between the semihard parton and the ridge hadron was carried out subsequently \cite{ch,hz1}.  The result can be stated in a succinct formula for a quantity called surface segment, $S(\phi,b)$, that can be expressed in terms of the elliptical integral of the second kind.  The mathematical details of $S(\phi,b)$ can be found in Ref. \cite{hz1}.  Only the physical meaning is given here.

In the approximation that the initial configuration of the dense medium created by a collision at $b$ is an ellipse in the transverse plane, $S(\phi,b)$ is the segment on the elliptical boundary through which semihard partons should be emitted if they are to contribute to the formation of any ridge particle that is directed at $\phi$.  The origin of such a segment function is that the conical region surrounding the trajectory of a semihard parton in which the medium partons can be thermally enhanced has a finite width $\sigma$, the value of which has previously been determined by fitting the correlation data \cite{ch}.  Combining that angular constraint with the property that hadrons are emitted, on average, normal to the surface in an expanding system, one can determine $S(\phi,b)$ unambiguously \cite{hz1}. Fig.\ 1(a) shows its $b$ dependence for $\phi=0$ and $\pi/2$.  It is the $\phi$ dependence of this function that specifies the azimuthal properties of the inclusive distributions of both pions and protons for any centrality.

 \begin{figure}[tbph]
\centering
\includegraphics[width=0.65\textwidth]{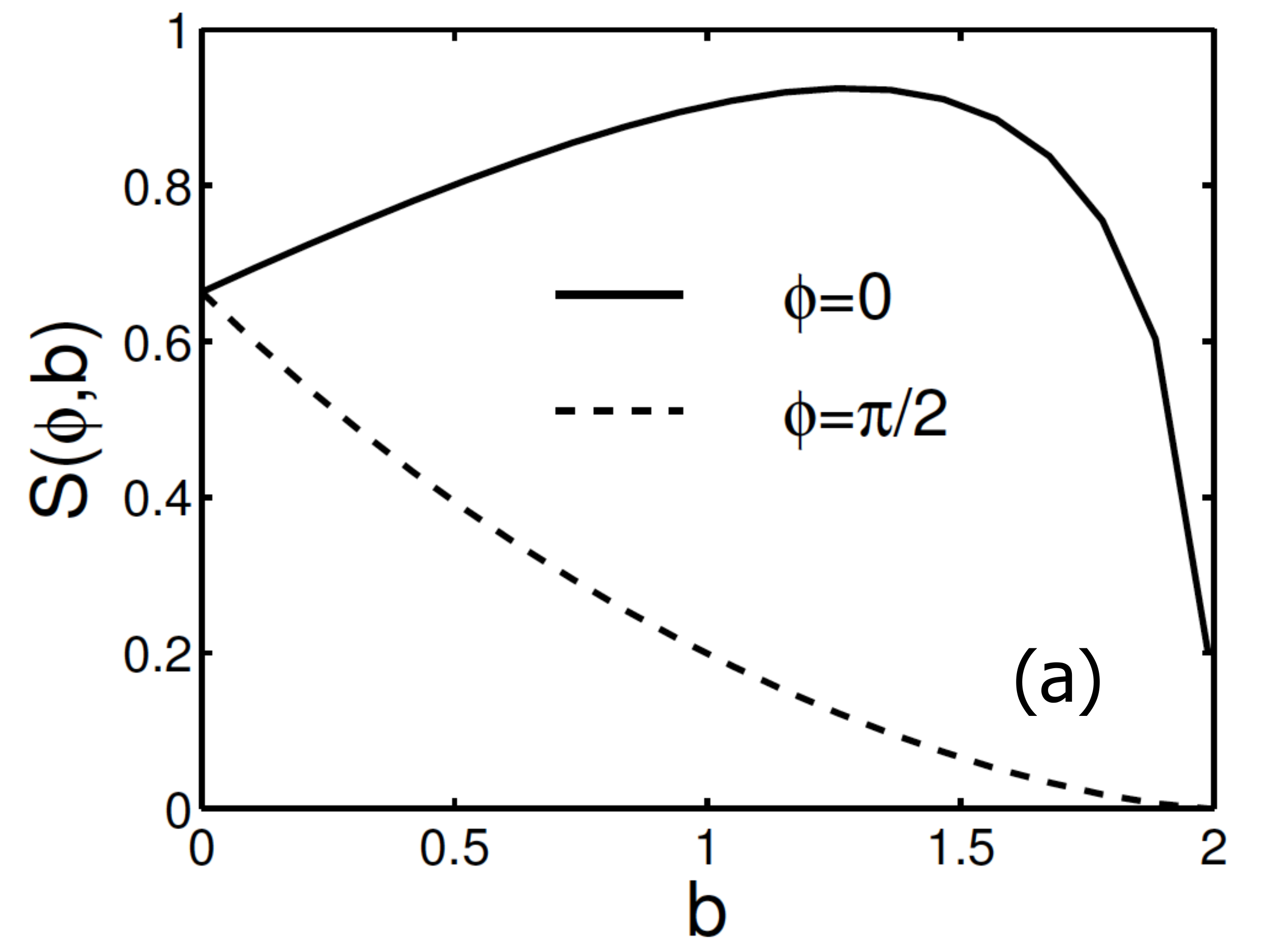}
\vspace*{-1cm}
\includegraphics[width=0.65\textwidth]{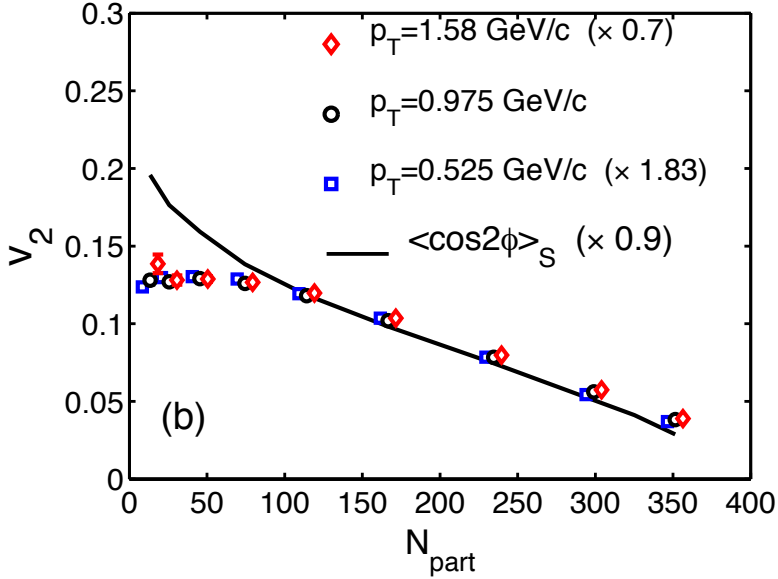}
\vspace*{-3cm}
\caption{(a) Surface segment $S(\phi,b)$ vs impact parameter normalized by $R_A$ for $\phi=0$ and $\pi/2$. See \cite{hz1}. (b) Common dependence of $v_2(p_T, b)$ on $N_{\rm part}$ for $p_T<2$ GeV/c with normalizations of data \cite{ja1} adjusted to show universal behavior. Solid line is $\left< \cos 2\phi\right>_S$ shifted down to show agreement with data at large $N_{\rm part}$ \cite{hz}.}
\end{figure}

Making explicit that $\phi$ dependence, the second term on the right side of (\ref{1}) can be written in the form
\begin{eqnarray}
R^h(p_T,\phi,b)=S(\phi,b) \bar R^h(p_T,b) ,  \label{2}
\end{eqnarray}
where $\bar R^h(p_T,b)$ is the $p_T$ distribution averaged over all $\phi$.  The second Fourier harmonic $v_2$ can now be calculated as follows:
\begin{eqnarray}
v_2^h(p_T,b) &=& \langle \cos 2\phi \rangle_{\rho}^h = {\int_0^{2\pi} d\phi \cos 2\phi\rho^h(p_T,\phi,b)\over \int_0^{2\pi} d\phi\rho^h(p_T,\phi,b)}   \nonumber \\     
&=& {\bar R^h(p_T,b){1\over 2\pi}\int_0^{2\pi} d\phi \cos2\phi S(\phi,b) \over B^h(p_T,b)+\bar R^h(p_T,b)} = {\langle \cos2\phi \rangle_S\over Z^{-1}(p_T) + 1},     \label{3}
\end{eqnarray}
where $Z(p_T)=\bar R^h(p_T)/B^h(p_T,b)$.  The numerator, $\langle \cos2\phi \rangle_S$, depends only on centrality and can be directly calculated independent of $p_T$.  The result is shown in Fig.\ 1(b), with appropriately chosen magnitude, to exhibit the agreement with the $N_{\rm part}$-dependence of $v_2$ for all $p_T < 2$ GeV/c and for $N_{\rm part} > 100$ \cite{hz}.  For very peripheral collisions the calculation of $S(\phi,b)$ is not reliable because semihard partons produced in very thin elliptical overlap can emerge from both sides of the not-very-dense medium.  Nevertheless, the results for $N_{\rm part} > 100$ reproduce the data so well without the use of any adjustable parameters that a strong case can be made to regard the minijet contribution as being important.  What remains is a demonstration that the $p_T$ dependence works as well.

Before the $p_T$ dependence of $v_2^h(p_T)$ is discussed, it is more appropriate to consider the $p_T$ dependence of the hadron spectra $\bar \rho^h(p_T,b)$ first.  If for $p_T < 2$ GeV/c it is written in the form
\begin{eqnarray}
\bar \rho^h(p_T,b) = {dN_h\over p_Tdp_T} = {\cal N}_h(p_T)e^{-p_T/T},     \label{4}
\end{eqnarray}
it can be shown in the recombination model (through TT and TTT recombination) that both pion and proton spectra can be expressed as in (\ref {4}) with a common exponential factor, the prefactor ${\cal N}_h(p_T)$ being the only difference between the two species due to difference in the wave functions of the hadrons in terms of valons ($x$ dependence of constituent quarks) \cite{hy}.  The value of $T$ is the same for thermal partons that recombine and for hadrons; it is found to be 0.283 GeV phenomenologically \cite{hz}.  Since hydrodynamics is not used in that study, the thermal distribution at the time of hadronization is not derived, but assumed.  If minijets are indeed important, it is not clear how a theoretical treatment that is based exclusively on hydro can yield a reliable value for $T$.  Minijets can produce soft partons that can become an important and indistinguishable part of the medium partons.  We have divided the inclusive hadron distributions into two parts in (\ref {1}), which cannot be directly verified by experiement.  However, the requirement that $B^h(p_T,b)$ has no $\phi$ dependence leads to the explicit expression for $v_2$ in (\ref {3}), thereby allowing the difference in the $p_T$ dependencies in $B^h(p_T,b)$ and $\bar R^h(p_T,b)$ to be determined indirectly. If they were the same, $v_2$ would have no $p_T$ dependence.

If we use the exponential form for $B^h(p_T,b)$
\begin{eqnarray}
B^h(p_T,b) = {\cal N}_h(p_T,b) e^{-p_T/T_0}     \label{5}
\end{eqnarray}
where $T_0$ is the background temperature, then $\bar R^h(p_T,b)$ has the form
\begin{eqnarray}
\bar R^h(p_T,b) = {\cal N}_h(p_T,b) \left[e^{-p_T/T}-e^{-p_T/T_0}\right],     \label{6}
\end{eqnarray}
so (\ref {3}) can be used to fit $v_2^h(p_T,b)$ with $T_0$ being the only adjustable parameter for both $h=\pi$ and $p$.  In Fig.\ 2 the STAR for 0-5\% centrality \cite{ja1} are shown, as fitted by $T_0 = 0.245$ GeV \cite{hz}.  Note that the normalization comes out right even though that is not adjustable, but fixed by (\ref {3}).  The fit is remarkable in that only one free parameter has been used.  Improvement can be made with TS recombination taken into account for a wider range of $p_T$, but more parameters will be necessary and more harmonics to be accounted for.

 \begin{figure}[tbph]
\centering
\vspace*{.5cm}
\includegraphics[width=0.4\textwidth]{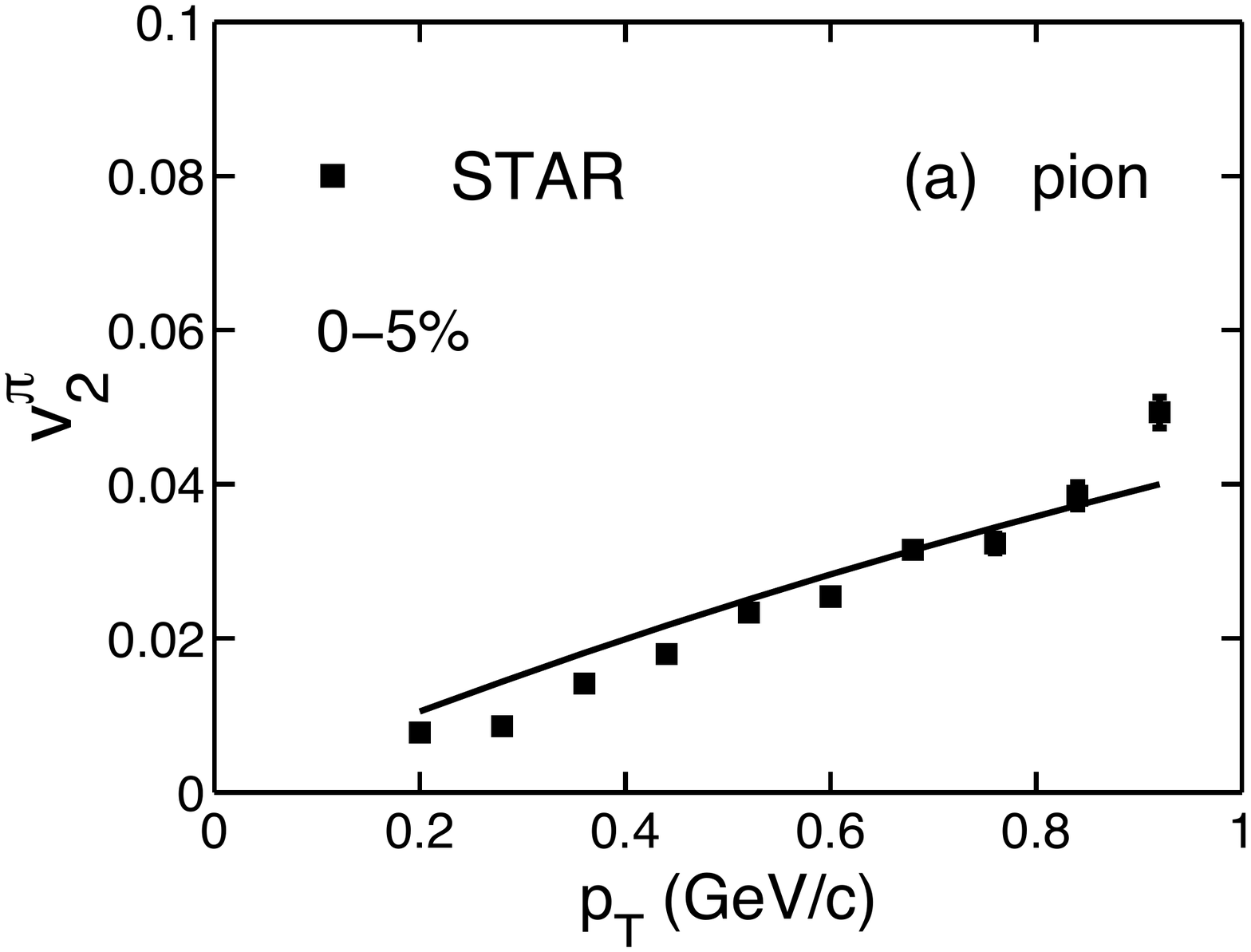}
\vspace*{-.5cm}
\hspace*{1cm}
\includegraphics[width=0.4\textwidth]{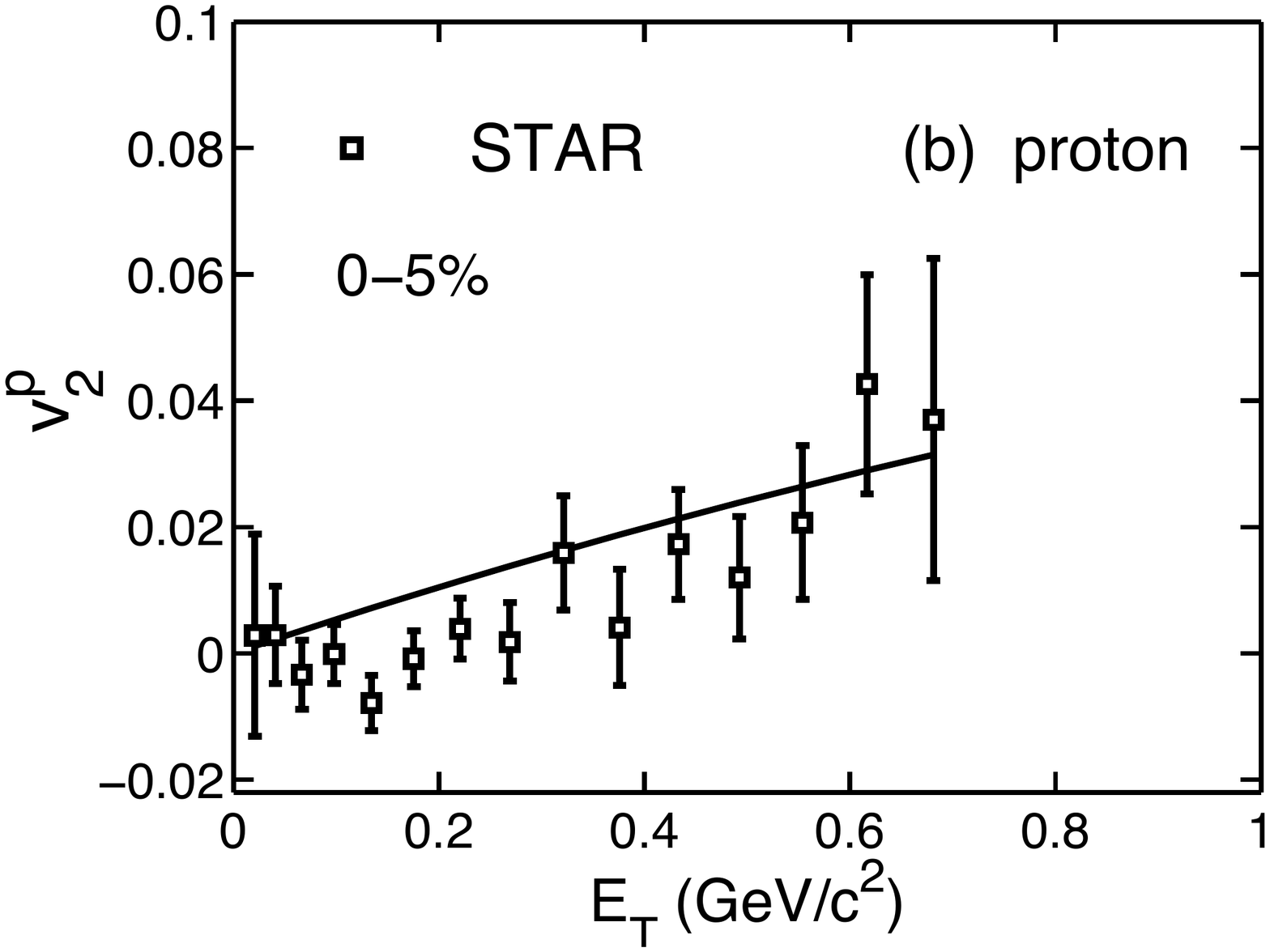}
\caption{$v_2$ for (a) pion and (b) proton at 0-5\% centrality. Data are from STAR \cite{ja1}, and solid lines from using Eq. (\ref{3}) using one parameter $T_0$ \cite{hz}.}
\end{figure}

It is evident from the above discussion that $v_2$ and ridge are intimately connected and that both are driven by minijets.  The $p_T$ distribution of the ridge, as expressed in (\ref {6}), can be approximated by ${\cal N}_h(p_T,b) \exp (-p_T/T_R)$, where $T_R$ turns out to 0.32 GeV.  Thus the ridge has a $T_R$ that is higher than the $T_0$ of the base.  That is the quantitative expression of the qualitative statement made earlier that the ridge contains enhanced thermal partons.  The result is from a study of the properties at $\eta \approx 0$.  The $\eta$ distribution of the ridge over a wide range involves the physics of the longitudinal motion of the initial partons, and has been discussed in Refs. \cite{ch1, ch2}, which will not be included in this brief review due to space limitation.

Fig.\ 2 shows only the very low $p_T$ region $< 1$ GeV/c.  To extend the region to $p_T \sim 2$ GeV/c requires the consideration of the TS component, since shower partons can contribute to low-$p_T$ partons.  Indeed, minijets can affect all higher harmonics.  In a sense minijets play a role similar to the fluctuations of initial eccentricity, except that they are actually fluctuations in momentum space and depend little on the details of flow dynamics.  We show in Fig.\ 3 only the result on $v_2$ after the TS recombination is taken into account \cite{hz}.  An additional parameter is used to specify the magnitude of the second harmonic of the minijet contribution; however, the $p_T$ and centrality dependencies are not adjustable.   The TS component is small for $p_T<1$ GeV/c, but significant in the interval $1<p_T<2$ GeV/c. The fits of the data from \cite{aa} in Fig.\ 3 are evidently excellent.

 \begin{figure}[tbph]
\centering
\vspace*{.5cm}
\includegraphics[width=0.6\textwidth]{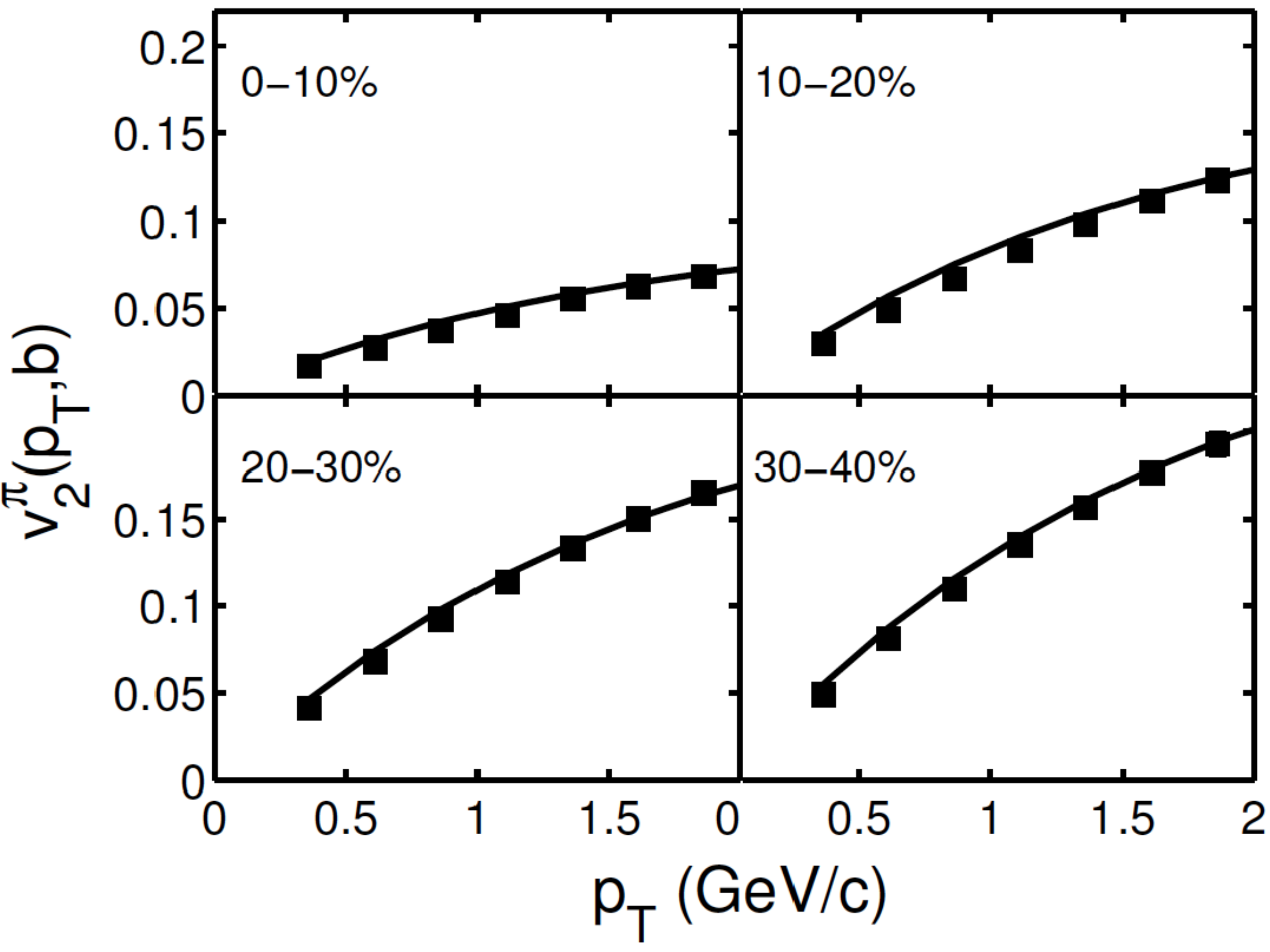}
\vspace*{-.5cm}
\caption{$v_2^\pi(p_T,b)$ for four centralities. Data are from Ref.\ \cite{aa}. The solid lines are from calculations in Ref.\ \cite{hz}.}
\end{figure}

\section{\large Conclusion}
The results summarized here demonstrate the importance of minijet contribution to the common observables in heavy-ion collisions.  If the possibility of fitting the data of those observables gives support to any particular approach, then the hydro and minijet approaches independently can claim sufficiency in explaining the data.  Neither are necessarily correct in all details.  Each emphasizes some aspect of the problem.  The hydro approach treats seriously the initial state and the evolutionary process of the fluid without considering minijets and the constituents of the hadrons formed.  The minijet approach is complementary.  At higher collision energies, such as at LHC, the probability of producing high densities of minijets is even higher.  To ignore their effects on the common observables, let alone hadrons correlated to jets, would seem to miss a large part of the whole story.  Our discussion here serves at least as a preliminary view of what is lacking in the so-called ``standard model", and may contribute to the broadening of the conventional wisdom.

\vskip 1cm

\centerline{\large\bf Acknowledgment}
Collaborations with L.\ Zhu, C.\ B.\ Yang and C.\ B.\ Chiu have been crucial in the work described, without whom no concrete results can be reported here. This writing was done with partial support  by the U.\ S.\ Department of Energy under Grant No. DE-FG02-96ER40972.


\end{document}